\documentclass[journal,letterpaper,twoside,final,twocolumn]{IEEEtran}%
\usepackage{amsmath}
\usepackage{amssymb}
\usepackage{amsfonts}
\usepackage{cite}
\usepackage{color}
\usepackage{comment}
\usepackage{graphicx}%
\newcommand{\rmd}{{\rm{d}}}

\DeclareMathOperator{\res}{res}
\DeclareMathOperator{\real}{Re}

\providecommand{\U}[1]{\protect\rule{.1in}{.1in}}
\providecommand{\U}[1]{\protect\rule{.1in}{.1in}}

\newtheorem{theorem}{Theorem}

\newtheorem{lemma}[theorem]{Lemma}

\allowdisplaybreaks[1]

\begin{document}

\title{A Theorem on the Asymptotic Outage Behavior of Fixed-Gain Amplify-and-Forward Relay Systems}

\author{Justin P. Coon, \IEEEmembership{Senior Member}
\thanks{J. P. Coon is with the Department of Engineering Science, University of Oxford, Parks Road, Oxford OX1 3PJ, U.K.; e-mail: justin.coon@eng.ox.ac.uk; tel: +44 (0)1865 273 027.}}

\maketitle

\begin{abstract}
A theorem that describes the high signal-to-noise ratio (SNR) outage behavior of fixed-gain amplify-and-forward (FGAF) relay systems is given.  Qualitatively, the theorem states that the outage probability decays according to a power law, where the power is dictated by the poles of the moments of the underlying per-hop fading distributions.  The power law decay is dampened by a logarithmic factor when the leading pole (furthest to the right in the plane) is of order two or more.  The theorem is easy to apply and several examples are presented to this effect.
\end{abstract}

\begin{IEEEkeywords}
Amplify-and-forward, outage, diversity.
\end{IEEEkeywords}

\section{Introduction}
Multi-hop fixed-gain amplify-and-forward (FGAF) relay systems have received a lot of attention recently due to their suitability for use in low-complexity systems, such as utility management applications and sensor networks \cite{Weber2009}.  The ability to analyze the performance of FGAF networks is of great importance, yet it can be difficult to obtain expressions for the outage probability of these systems.  Although some exact results exist for simple two-hop networks (see, e.g.,~\cite{DiRenzo2009}), it is often necessary to resort to high signal-to-noise ratio (SNR) approximations or diversity analysis to glean some insight to more complicated systems~\cite{Xu2010,Ding2010}.  Such endeavors are frequently undertaken for specific classes of channel model, such as Nakagami-$m$ fading~\cite{Karagiannidis2006} or the case where the per-hop channel power densities are non-zero at the origin~\cite{Farhadi2008}.

In this letter, we seek to somewhat unify and extend the excellent advancements made in, for example,~\cite{Karagiannidis2006,Farhadi2008} by providing a simple framework that can be used to ascertain, at least partially, the outage behavior\footnote{By \emph{behavior} we mean the outage decay rate at high SNR.} of FGAF systems operating in more general scenarios, including inhomogeneous fading conditions. 
To this end, we detail a theorem that describes how the moments of the underlying per-hop fading distributions completely characterize the high SNR outage behavior for a broad class of channel models.  The theorem can easily be employed in numerous settings as we will show through several examples in the sections that follow.

\section{System Model}

Consider an $N$-hop network with a source node, a destination node, and $N-1$ half-duplex relay nodes in between. A data symbol $d$, with $\mathbb{E}[|d|^{2}]=1$, is conveyed from the source to the first relay node. This symbol is affected by a flat fading channel and additive Gaussian noise at the receiver. The received signal is then amplified by a fixed gain $\alpha_{1}$, then conveyed to the next relay node and so on until the destination is reached. Let $h_{n}$ be the channel coefficient for the $n$th hop. Also, denote the additive noise at the $n$th relay node (or the destination) by $v_{n}$, which is zero-mean complex Gaussian distributed with variance $\sigma_{n}^2/2$ per dimension. Now, we can write the signal received at the destination as
\begin{equation}
r=\left(  \prod_{n=1}^{N}\alpha_{n-1}h_{n}\right)  d+\sum_{n=1}^{N}v_{n}\!\!
\prod_{j=n+1}^{N}\!\!\alpha_{j-1}h_{j}   \label{eq:sys_model}%
\end{equation}
where $\alpha_{0}=1$. In order to maintain generality, we do
not explicitly define $\alpha_{n}$ except to note that it is a constant.%


Define the parameter $\bar{\gamma}$ to be the average \emph{transmit} SNR at the source, and let $\bar{\gamma}={\rho_{n}}/\sigma_{n}^2$ for $n=1,\ldots,N$ where $\{\rho_{n}\}$ are strictly positive (and finite) scaling factors and, clearly, $\rho_{1}=1$. Letting $X_{n}=\alpha_{n-1}^{2}|h_{n}|^{2}$, we can now write the instantaneous end-to-end
SNR for the $N$-hop link as%
\begin{equation}
{\sf{SNR}}=\frac{\prod_{n=1}^{N}X_{n}}{\sum_{n=1}^{N}\rho_{n}\prod
_{j=n+1}^{N}X_{j}}\bar{\gamma}. \label{eq:snr}%
\end{equation}
Given an SNR outage threshold $\gamma_{t}$, we define the outage probability
to be
\begin{equation}
  p_{o}(\bar{\gamma})= \mathbb P({\sf{SNR}}(\bar{\gamma})<\gamma_{t}).
\end{equation}

\section{Outage Behavior for large $\bar\gamma$}
We now state a theorem that describes the outage behavior of a diverse class of FGAF networks for large $\bar\gamma$.
\begin{theorem}\label{th:1}
Let $X_1,\ldots,X_N$ be a set of independent random channel gains as described above with $f_{X_n}$ denoting the probability density function (p.d.f.) of $X_n$.  Assume all positive moments $\mathbb E[X_n^{a}]$ exist for $1\leq a < \infty$, and define the product
\begin{equation}\label{eq:G(s)}
  G(s) = \prod_{n=1}^N \mathbb E[X_n^s]
\end{equation}
for some $s\in\mathbb C$.  Suppose that $G(s)$ exhibits a $k$th order pole at $s=s_0$, and that no other pole of $G(s)$ lies to the right (in the real sense) of this point.  If, for some $b>0$, $G(s)$ vanishes like $O(|s|^{-b|s|})$ in the left half $s$-plane as $|s| \rightarrow \infty$, then the outage probability decays asymptotically like
\begin{equation}
  p_o(\bar\gamma)= O((\ln \bar\gamma)^{k-1} \bar\gamma^{\real(s_0)}), \quad \bar\gamma \rightarrow \infty.
\end{equation}
\end{theorem}
\begin{IEEEproof}
  See the appendix.
\end{IEEEproof}

Theorem \ref{th:1} deserves some discussion.  The product of moments in (\ref{eq:G(s)}) arises from the fact that the information component of the signal reaches the destination having been affected by the product of independent random channel states (cf.~(\ref{eq:snr})).  Although the theorem caters for complex valued poles of $G(s)$, the underlying distributions in practice are real-valued, and thus $s_0 \in\mathbb R$ in a typical example.  The finite-moment condition stated in the theorem hypothesis is not as restrictive as it may first appear.  We will see that many practical fading distributions adhere to this condition.  As a general rule, distributions that belong to the exponential family satisfy the hypothesis, which includes most small-scale fading models; heavy-tailed distributions do not obey the hypothesis\footnote{Log-normal fading is one example: $|G(s)|$ diverges in the left half-plane.}.

A direct corollary of Theorem \ref{th:1} is that the diversity order, defined in the usual way, is given by
\begin{equation}\label{eq:diversity}
  d = \lim_{\bar\gamma \rightarrow \infty}\frac{\ln p_o(\bar\gamma)}{-\ln\bar\gamma} = -s_0
\end{equation}
where, from here on, we assume that $s_0\in\mathbb R$ as discussed above.  At this point, our physical understanding of the system suggests that $s_0 < 0$ for practical fading distributions.  Indeed, this is the case as we will see below.  

For large but finite $\bar\gamma$, Theorem~\ref{th:1} shows that the rate of convergence to the asymptote is diminished by the order $k$ of the pole at $s_0$ and the logarithm of $\bar\gamma$.  Specifically, the finite diversity order can be written as
\begin{equation}\label{eq:finite_diversity}
  d(\bar\gamma) = -s_0 - (k-1)\frac{\ln\ln\bar\gamma}{\ln\bar\gamma} + O(1/\ln\bar\gamma).
\end{equation}
In contrast, direct links typically converge like $O(1/\ln\bar\gamma)$.  


\section{Applications}\label{sec:applications}
In this section, we apply Theorem \ref{th:1} to study the asymptotic outage behavior of several practical systems.  We begin with straightforward multi-hop homogeneous scenarios\footnote{Here, a \emph{homogeneous} relay network is one for which the fading processes at each hop adhere to the same distribution, but are not necessarily identical.}.  We then briefly discuss the more complex case of inhomogeneous fading.  Where possible, we compare our results with previously published results.

\subsection{Nakagami-$m$ and Weibull}\label{sec:nakagami}

For the $N$-hop homogeneous network considered here, denote the scale and shape parameters related to the distribution of the gain variable $X_n$ corresponding to the $n$th hop by $\theta_n$ and $m_n$, respectively.  The p.d.f. of $X_n$ can be written as
\begin{equation}\label{eq:pdf_nakagami}
  f_{X_n}\!(x)=\frac{\omega_n}{\theta_n^{m_n}\nu_n}x^{m_n-1}e^{-(x/\theta_n)^{\omega_n}},\quad x\geq 0 
\end{equation}
where $\omega_n = 1$ and $\nu_n = \Gamma(m_n)$ if $X_n$ is Nakagami-$m$ distributed, and $\omega_n = m_n$ and $\nu_n = 1$ if it is Weibull distributed\footnote{Note that this scenario cannot be treated with the theory detailed in~\cite{Farhadi2008} since $m_n > 1$ is possible, in which case the p.d.f.s are zero at the origin.}.  The $s$th moment is given by
\begin{equation}\label{eq:moment_nakagami}
  \mathbb E[X_n^s] = \theta_n^{s}\Gamma((s+m_n)/\omega_n)/\nu_n, \quad \real(s)>-m_n.
\end{equation}
On inspection, we see that the hypothesis of Theorem \ref{th:1} is satisfied\footnote{The moments are clearly bounded for $1 \leq \real(s) < \infty$, and the rate of decay of $G(s)$ in the left half-plane follows from Stirling's formula.}, and that $G(s)$ has poles at $s = -m_n - \omega_n j$ for $j = 0,1,\ldots$ and $n = 1,\ldots,N$.  Since $m_n\in\mathbb R$, the poles are real, and the pole that lies furthest to the right is located at $\max_n\{-m_n\}$. Hence, this pole becomes $s_0$ in Theorem \ref{th:1}.  If $k\leq N$ channels are parameterized by the minimum shape factor, then the order of this pole is $k$ and
\begin{equation}
  p_o(\bar\gamma) = O((\ln\bar\gamma)^{k-1}/\bar\gamma^{\min_n\{m_n\}}),\quad\bar\gamma\rightarrow\infty.
\end{equation}
The corresponding finite diversity expression and limit are
\begin{equation}
  d(\bar\gamma) = \min_n\{m_n\} - (k - 1)\frac{\ln\ln\bar\gamma}{\ln\bar\gamma} + O(1/\ln\bar\gamma).
\end{equation}
and $ d = \min_n\{m_n\}$. For $N=2$, it is easy to see that this result agrees with that given for Nakagami-$m$ channels in~\cite{Xu2010} for a two-hop system and indeed generalizes that contribution to Weibull channels and multi-hop networks in a natural way.  

\subsection{Rician and Hoyt}
Homogeneous Rician and Hoyt (Nakagami-$q$) fading channels yield similar asymptotic behavior.  For Rician fading, $X_n$ is a noncentral $\chi^2$ random variable, and we have
\begin{equation}
  f_{X_n}\!(x) = \frac{K_n+1}{\theta_n e^{K_n}}  e^{-\frac{K_n+1}{\theta_n}x} I_{0}\!\left(\sqrt{\frac{4K_n\left(  K_n+1\right)x}{\theta_n}}\right),\, x\geq 0
\end{equation}
where $\theta_n$ and $K_n$ are the scale and $K$ factors, respectively, and $I_0(\cdot)$ is the zeroth-order modified Bessel function of the first kind.  The $s$th moment can be calculated to be
\begin{multline}
  \mathbb E[X_n^s] = e^{-K_n}\!\left(  \frac{\theta_n}{K_n+1}\right)^{s}\Gamma( s+1)  \,_{1}F_{1}(  s+1,1;K_n),\\
  \real(s) > -1
\end{multline}
where$\,\,_{1}F_{1}(\cdot,\cdot;\cdot)$ is the confluent hypergeometric function, which is analytic in its first argument.  Thus, $G(s)$ has an $N$th-order pole at $s = -1-j$ for $j = 0,1,\ldots$, and the asymptotic outage behavior follows $p_o(\bar\gamma) = O((\ln\bar\gamma)^{N-1}/\bar\gamma)$.

For homogeneous Hoyt fading, we have
\begin{equation}
  f_{X_n}\!(x) = \frac{1+q_n^{2}}{2q_n\theta_n}e^{-\frac{\left(  1+q_n^{2}\right)  ^{2}}{4q_n^{2}\theta_n}x}I_{0}\!\left(  \frac{1-q_n^{4}}{4q_n^{2}\theta_n}x\right),\quad x\geq 0 
\end{equation}
and
\begin{multline}
  \mathbb E[X_n^s] = \left(  \frac{2q_n}{1+q_n^{2}}\right)  ^{2s+1}\theta_n^{s}\Gamma\left(  s+1\right)\\
  \times\,_{2}F_{1}\!\left(  \frac{s+1}{2},\frac{s+2}{2};1;\left(  \frac{1-q_n^{2}}{1+q_n^{2}}\right)  ^{2}\right),\,\real(s) > -1
\end{multline}
where$\,\,_{2}F_{1}(\cdot,\cdot;\cdot;\cdot)$ is the hypergeometric function, which is analytic in its first two arguments.  The pole structure of $G(s)$ is the same as for the Rician case, and a similar result follows.

\subsection{Inhomogeneous Channels}
For FGAF networks experiencing inhomogeneous fading channels, the asymptotic behavior is determined by the channel that has an $s$th moment with dominant corresponding pole.  Thus, if we consider a system with $N = 3$ hops adhering to a Rician/Nakagami-$m$/Weibull fading configuration, where the shape parameters related to the second and third hops are greater than one, the diversity order is limited to one through the Rician channel, and $p_o(\bar\gamma) = O(1/\bar\gamma)$ as $\bar\gamma\rightarrow\infty$.  Reordering the channel configuration does not change the asymptotic slope of $p_o(\bar\gamma)$, but it does affect the coding gain\footnote{We do not directly treat the coding gain in this general framework, primarily because it tends to be particular to a given fading configuration; however, the coding gain can be inferred from the asymptotic analysis detailed in the appendix.}.  This points to the notion of an \emph{asymptotic bottleneck}; the channel(s) exhibiting the least favorable diversity properties will govern asymptotic performance, and multiple occurrences of such pseudo-identically distributed channels -- \emph{pseudo-}identical in the sense that shape parameters may be the same, but scale parameters may differ -- will cause convergence to the asymptote to occur at a rate less than $O(1/\ln\bar\gamma)$.

\section{Numerical Results}\label{sec:results}
\begin{figure}[t]
\begin{center}
\includegraphics[width=8cm]{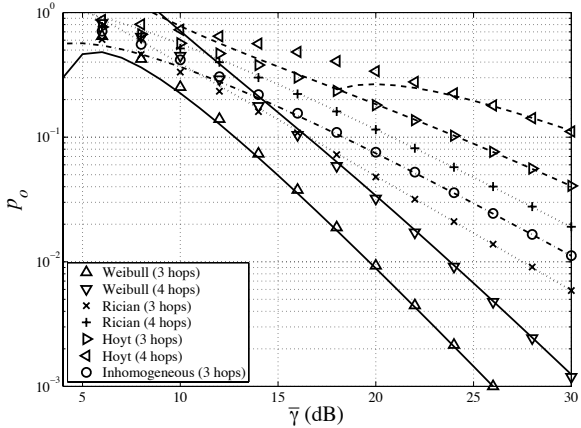}
\caption{Outage probability vs. $\bar\gamma$ for $N=3,4$ with $\theta_n = \rho_n = 1$ for all $n$.  For the homogeneous systems, $m_n = 2.2,1.8,1.8,1.8$; $K_n = 1,3,5,0$; and $q_n = 3/4,1/2,1/3,1/4$.  For the inhomogeneous system, $m_1 = 2$, $K_2 = 1$, and $q_3 = 1/2$.  Markers indicate simulation results and lines represent theory.}
\label{fig:1}
\end{center}
\end{figure}

In Fig. \ref{fig:1}, the theory detailed above is compared to numerical simulations for three and four hop Weibull, Rician, and Hoyt channel models.  Additionally, a three hop inhomogeneous example is illustrated, which consists of a Weibull distributed first hop channel followed by a Rician distributed second hop and finally a Hoyt distributed third hop.  The distribution parameters are given in the figure caption.  To generate the theoretical curves, the leading order residues of the asymptotic expansion were calculated (see (\ref{eq:Po_final}) and (\ref{eq:Ipsi}) in the appendix).  This comparison clearly validates the asymptotic theory; indeed, for many cases, the asymptotic expressions are accurate even at moderate values of $\bar\gamma$.  The non-monotonic behavior of the three hop Weibull and four hop Hoyt theoretical curves is a result of the higher-order pole structure of $G(s)$, which implies the leading order expression for $p_o$ contains a logarithmic term in the numerator (cf. Theorem \ref{th:1}).  In fact, although it is not readily apparent from the curves, this is true of all of the examples explored in Fig. \ref{fig:1}.

\section{Conclusions}\label{sec:conclusions}
In this letter, it was shown that the outage probability of an FGAF system decays according to a power law, where the power is dictated by the poles of the moments of the underlying fading distributions.  The power law decay is dampened by a logarithmic factor when the leading pole is of order two or more.  The theory detailed herein was applied to several practical examples, which both corroborated previously reported results and extended them to more complex scenarios.

\appendix[Proof of Theorem 1]
We first state and prove the following useful lemma.

\begin{lemma}\label{lem:1}
Let $f:[0,\infty)\mapsto\lbrack0,\infty)$ be a smooth function (i.e., all derivatives exist) of a single variable where $\lim_{x\rightarrow0,\infty}x^{s-r-1}f^{(\ell-r-1)}(x)=0$ for $ r=0,\ldots,\ell-1$.  Let $F(s)=\int_{0}^{\infty}x^{s-1}f(x)\rmd x$ denote the Mellin transform of $f$.  Define $f_{\epsilon}(x):= f(x+\epsilon)$ for $\epsilon>0$. Then, the Mellin transform of $f_{\epsilon}$ has the asymptotic formal power series representation
\begin{equation}\label{eq:Mfeps}
 F_\epsilon(s)=\sum_{\ell=0}^{\infty}(-1)^{\ell}\frac{\Gamma(s)}{\Gamma(s-\ell)}\frac{F(s-\ell)}{\ell!}\epsilon^{\ell},\quad\epsilon\rightarrow 0.
\end{equation}
\end{lemma}

\begin{IEEEproof}
Expanding $f_{\epsilon}$ near $x$ gives $F_{\epsilon}(s)=\int_{0}^{\infty}x^{s-1}( \sum_{\ell=0}^{\infty}f^{(\ell)}(x)\epsilon^{\ell}/\ell!) \, \rmd x$. 
In general, the series diverges. However, we require a \emph{formal} power series~\cite{Niven1969}, and thus we integrate term-by-term to obtain (\ref{eq:Mfeps}), applying~\cite[eq. (3.1.9)]{Paris2001} in the process. The sequence $\{\epsilon^{\ell}\}$ is an asymptotic scale; hence, (\ref{eq:Mfeps}) is a Poincar\'e expansion~\cite{Paris2001}.
\end{IEEEproof}

We now proceed with the proof of Theorem~\ref{th:1}.  From (\ref{eq:snr}), the outage probability can be written as
\begin{multline}
\mathbb P\!\left(  \prod_{n=1}^{N}X_{n}-\sum_{n=1}^{N-1}\xi_{n}\prod_{j=n+1}^{N}%
X_{j}<\xi_{N}\right)  =\\
1 - \mathbb P((\cdots((X_{1}-\xi_{1})X_{2}-\xi_{2})\cdots)X_{N} \geq \xi_{N})
\label{eq:Po_telescope}%
\end{multline}
where $\xi_{n}(\bar{\gamma})={\rho_{n}\gamma_{t}}/{\bar{\gamma}}$.  Define the auxiliary random variables
\begin{equation}\label{eq:Zn}
Z_{n}=W_{n}X_{n+1} \quad\text{and}\quad W_{n}=Z_{n-1}-\xi_{n}>0
\end{equation}
for $n=0,\ldots,N-1$ with $W_{0}=1$. Note that $W_{n}$ is a conditional random variable in that it relates to the translation of $Z_{n-1}$ where it is given that $Z_{n-1}>\xi_{n}$. Also, $W_{n}$ and $X_{n+1}$ are independent. Applying this same conditioning on (\ref{eq:Po_telescope}) recursively, we can rewrite the outage probability as
\begin{equation}\label{eq:PZn}
  p_{o}=1-\prod_{n=1}^{N}\mathbb P(Z_{n-1}>\xi_{n}). 
\end{equation}

We require an expression for $\bar p_{Z_{N-1}}\!(\xi_{N}):= \mathbb P(Z_{N-1}>\xi_{N})$ in order to compute the outage probability.  We can write the Mellin transform of $\bar p_{Z_{N-1}}$ as
\begin{align}\label{eq:MFZN-1}
  \bar P_{Z_{N-1}}\!(s)  &=\int_{0}^{\infty}x^{s-1}\bar p_{Z_{N-1}}\!(x)\rmd x \nonumber\\ 
  &=s^{-1}F_{Z_{N-1}}\!(s+1),\quad\operatorname{Re}(s)>0 
\end{align}
where $F_{Z_{N-1}}$ is the Mellin transform of the p.d.f. $f_{Z_{N-1}}$ of $Z_{N-1}$, and the second equality follows from integration by parts.  Upon taking the inverse transform (cf. \cite[eq. 6.1 (1)]{Erdelyi1954}) of (\ref{eq:MFZN-1}) and substituting into (\ref{eq:PZn}), we can write
\begin{equation}\label{eq:Po}
  p_{o}=1-\frac{\Omega_{N-1}}{2\pi i}\int_{L}\xi_{N}^{-s}s^{-1} F_{Z_{N-1}}\!(s+1)\rmd s
\end{equation}
where $\Omega_{N-1}=\prod_{j=1}^{N-1}\bar p_{Z_{j-1}}\!(\xi_{j})$.  The path of integration is the line $L=\{c+it:-\infty < t < \infty\}$ for some $c\in\mathcal{A}$ with $\mathcal{A}$ being the (nonempty) strip of analyticity of $s^{-1}F_{Z_{N-1}}\!(s+1)$.

Eq.~\eqref{eq:Po} presents a problem in that it depends upon $\Omega_{N-1}$ and the transform $F_{Z_{N-1}}$, neither of which possesses a simple representation.  Thus, we require the following result.

\begin{lemma}\label{lem:2}
The Mellin transform $F_{Z_{n}}$ admits the asymptotic formal power series representation
\begin{multline}
  F_{Z_{n}}\!(s)=\frac{\Gamma(s)}{\Omega_{n}}\sum_{\lambda_N=0}^{\infty}\frac{(-{\gamma_{t}})^{\lambda_N}}{\Gamma(s-\lambda_N)\bar{\gamma}^{\lambda_N}}\!\!\!\!\sum_{\substack{\ell_{1},\ldots,\ell_{n}:\\\ell_{1}+\cdots+\ell_{n}=\lambda_N}}\prod_{j=1}^{n}\frac{{\rho_{j}^{\ell_{j}}}}{\ell_{j}!}\\
\times\prod_{j=1}^{n+1}F_{X_{j}}\!(s-\lambda_N+\lambda_{j}),\quad\bar{\gamma}\rightarrow\infty
\end{multline}
for $n=1,2,\ldots\,$, where $\lambda_{j}={\textstyle\sum\nolimits_{r=1}^{j-1}}\ell_{r}$ with $\lambda_1 = 0$, and the second sum is over all weak compositions of $\lambda_N$ in $n$ parts.
\end{lemma}

\begin{IEEEproof}
First, we note that $F_{Z_{0}}\!(s)  =F_{X_{1}}\!(s)$. From~(\ref{eq:Zn}) and properties of Mellin transforms (cf. \cite{Erdelyi1954}), we have that $F_{Z_{n}}\!(s)  =F_{W_{n}}\!(s) F_{X_{n+1}}\!(s)  $. It is easy to see from the definition of $W_{n}$ that its p.d.f. is given by $f_{W_{n}}\!(  w)  =f_{Z_{n-1}}\!(w+\xi_{n})  /\bar p_{Z_{n-1}}\!(\xi_{n})  $ for $w\geq0$, and thus one can verify that the lemma holds for $n=1$ by using Lemma~\ref{lem:1}. The proof for $n>1$ follows from induction and manipulations of the formal power series~\cite{Niven1969}, where~\cite[6.1 (10)]{Erdelyi1954} and Lemma~\ref{lem:1} are applied in the inductive step.
\end{IEEEproof}

Now, we can apply Lemma~\ref{lem:2} to~\eqref{eq:Po} with $n = N-1$, integrate term by term, substitute $s \leftarrow s - \lambda_N$, and rearrange the result to obtain the formal series
\begin{equation}\label{eq:Po_final}
  p_{o} = 1-\!\sum_{\lambda_N=0}^\infty \sum_{\substack{\ell_{1},\ldots,\ell_{N-1}:\\\ell_{1}+\cdots+\ell_{N-1}=\lambda_N}}
    \!\!\prod_{j=1}^{N-1}\frac{(-\rho_j)^{\ell_j}}{\rho_N^{\ell_j}\ell_j!} I(\xi_N;\ell_1,\ldots,\ell_{N-1})
\end{equation}
for $\bar{\gamma}\rightarrow\infty$, where the integral $I(\xi_{N};\ell_1,\ldots,\ell_{N-1})$, which we denote by $I$ for brevity, is given by
\begin{equation}\label{eq:Ipsi}
  I=\frac 1 {2\pi i} \int_{L'} \xi_N^{-s}\frac{\Gamma(s+\lambda_N)\prod_{j=1}^N F_{X_j}\!(s+1+\lambda_j)}{\Gamma(s+1)} \rmd s.
\end{equation}
The path of integration is the line $L' = \{c-\lambda_N+it : -\infty < t < \infty \}$ with $c$ lying to the right of the poles of the integrand.

The $\lambda_N = 0$ term in the outer summation of (\ref{eq:Po_final}) is
\begin{align}\label{eq:po_expanded_1st}
  \frac 1 {2\pi i} \int_{c-i\infty}^{c+i\infty} \frac{\xi_N^{-s}}{s}\prod_{j=1}^N F_{X_j}\!(s+1) \rmd s
\end{align}
the integrand of which, denoted herein by $T(s)$, has a simple pole at $s=0$.  Since $\lambda_N \geq 1$ for all other terms in (\ref{eq:Po_final}), and since all moments $\mathbb E[X_j^s] = F_{X_j}\!(s+1)$ exist by hypothesis, the integrands in those terms are well behaved at $s=0$.  Again, by hypothesis, the product $\prod_{j=1}^N F_{X_j}\!(s+1)$ vanishes sufficiently quickly as $|s|\rightarrow\infty$ in the left half-plane.  Thus, we can close the path of integration for (\ref{eq:po_expanded_1st}) with a semi-circle traversed counter-clockwise through the left half-plane from $+i\infty$ to $-i\infty$ such that it does not pass through a pole of the integrand.  The residue theorem can then be invoked to evaluate the integral, the result being the sum of the residues of the integrand.  The residue at $s=0$ is given by
\begin{equation}
  \res\{T,0\} = \lim_{s\rightarrow 0} s\, T(s) = \prod_{j=1}^N F_{X_j}\!(1) = 1
\end{equation}
which cancels the leading ``$1$'' in (\ref{eq:Po_final}).  

The remaining residues of $T(s)$ come from the product $G(s) = \prod_{j=1}^N F_{X_j}\!(s+1)$.  Now suppose $T(s)$ has a $k$th order pole at $s=s_0$, and there are no poles to the right of this in the $s$-plane.  Letting $H(s) = s^{-1}(s-s_0)^k G(s)$, the residue at $s=s_0$ can be evaluated by the limit
\begin{align}
  \res\{T,s_0\} &= \frac{1}{\Gamma(k)}\lim_{s\rightarrow s_0} \frac{\partial^{k-1}}{\partial s^{k-1}}\{(s-s_0)^{k}\,T(s)\} \nonumber \\
  &= \frac{1}{\Gamma(k)} \sum_{m=0}^{k-1}\binom{k-1}{m}H^{(k-1-m)}(s_0)\frac{(-\ln\xi_N)^m }{\xi_N^{s_0}}
\end{align}
where $H^{(n)}(s_0)$ is the $n$th derivative of $H$ evaluated at $s\rightarrow s_0$.  Recall that $\xi_N = \gamma_t \rho_N/\bar\gamma$.  It follows that
\begin{equation}
  \res\{T,s_0\} = \frac{H(s_0)}{(\gamma_t \rho_N)^{s_0} \Gamma(k)} (\ln\bar\gamma)^{k-1} \bar\gamma^{s_0} + o((\ln\bar\gamma)^{k-1} \bar\gamma^{s_0}).
\end{equation}
One can confirm by inspection of the $\lambda_N > 0$ integrals in~\eqref{eq:Po_final} that no poles lie to the right of $s_0$ since any addition of a positive number to an argument of $F_{X_j}$ will shift the locations of the poles to the left.  Thus, the pole at $s=s_0$ is the first pole (scanning right to left), apart from the simple pole at the origin already accounted for, exhibited by \emph{any} integrand in the expression for $p_o$.  The theorem is proved by noting that only asymptotic behavior is sought, and thus the convergence of (\ref{eq:Po_final}) is neither guaranteed nor required.

\bibliographystyle{IEEEtran}
\bibliography{acompat,IEEEabrv,multihop_relay_analysis_bibfile}

\end{document}